\documentclass[final,5p,times,twocolumn,authoryear]{elsarticle}

\usepackage{amsmath}
\usepackage{amssymb}
\usepackage{booktabs}
\usepackage{url}
\usepackage{hyperref}

\newcommand{\paperfigure}[2]{%
  \IfFileExists{#2}{%
    \includegraphics[width=#1]{#2}%
  }{%
    \PackageWarningNoLine{init-angles}{Figure file missing: \detokenize{#2}}%
    \fbox{\parbox[c][\dimexpr#1*2/5\relax][c]{\dimexpr#1-2\fboxsep-2\fboxrule\relax}{%
      \centering\small Figure file missing\par
      \scriptsize\path{#2}}}%
  }%
}

\journal{High Energy Astrophysics}

\begin{document}
\begin{frontmatter}

\title{Initial magnetic obliquities of normal radio pulsars from their kinematic ages}

\author[first]{Ekaterina Tokareva\corref{cor1}}
\cortext[cor1]{Corresponding author}
\ead{e.tokareva135@gmail.com}
\author[second,third]{Anton Biryukov}
\ead{ant.biryukov@gmail.com}

\affiliation[first]{organization={Moscow Institute of Physics and Technology},
            addressline={Institutsky per. 9},
            city={Dolgoprudny},
            state={Moscow Region},
            postcode={141700},
            country={Russia}}
\affiliation[second]{organization={The Raymond and Beverly Sackler School of Physics and Astronomy, Tel Aviv University},
            addressline={55 Chaim Levanon St.}, 
            city={Tel Aviv},
            postcode={6997801},
            country={Israel}}
\affiliation[third]{organization={Sternberg Astronomical Institute, Lomonosov Moscow State University},
            addressline={13 Universitetskij pr.}, 
            city={Moscow},
            postcode={119234}, 
            country={Russia}}

\begin{abstract}
We study 35 isolated radio pulsars with known kinematic ages in the
range $\sim$0.2--60~Myr and published estimates of the magnetic angle (obliquity)
$\alpha$ between the magnetic and spin axes. Using the conventional model of pulsar spin-down, which predicts a
gradual alignment of the magnetic and spin axes, and a phenomenological
law of magnetic field decay, we integrate the evolution of each pulsar
backward in time and estimate its initial magnetic angle $\alpha_0$
and initial spin period $P_0$ self-consistently. The uncertainties of the ages and of
the magnetic angles are propagated by a Monte Carlo procedure, and
the geometric (beaming) correction is applied. 
These are the first estimates of $\alpha_0$ based on ages that are independent of
the pulsar timing. The reconstructed distribution of $\alpha_0$ is
concentrated at intermediate angles and shows a lack of pulsars with
$\alpha_0 \gtrsim 70^\circ$ with respect to both the isotropic distribution,
$p(\alpha_0) \propto \sin\alpha_0$, and the distribution uniform in
angle. Tests with mock pulsar samples show that the method recovers
individual values of $\alpha_0$ reasonably well, but that
part of the deficit can be produced by the method itself. The reconstructed initial periods, with a
mean of $0.4$--$0.5$~s, are formally longer than those found in earlier
works. Still, the same tests show that the method overestimates $P_0$,
so that this result cannot be taken at face value.
\end{abstract}


\end{frontmatter}

\section{Introduction}
\label{sec:introduction}

Classical, rotation-powered radio pulsars are relatively young, isolated
neutron stars observed at the so-called ejector stage. Their magneto-
rotational evolution is essentially decoupled from the properties of the
surrounding interstellar medium and is governed only by the physics
inside their magnetospheres and interiors~\citep{2024Galax..12....7A}.

The major observable characterizing the evolution of a pulsar is
its spin period $P$. It is measured with high precision and is
now known for more than $3500$ objects~\citep{atnf}. A neutron star, affected by magnetospheric torques, gradually loses rotational energy and spins down at rate $\dot P$. The latter depends on the period itself, as well as on the stellar magnetic moment $\mu$ and on its
orientation with respect to the rotation axis~\citep[see e.g.][]{spitkovsky06, phil14, tche13, phil15}.

The orientation of the star dipole magnetic field is quantified by the so-called magnetic angle, $\alpha$, defined as the angle between the spin and magnetic axes. In principle, $\alpha$ is also an observable quantity. A variety of methods for estimating it have been proposed, and several thousand such estimates have been obtained for hundreds of objects \citep[e.g.][and references therein]{nikitina11, malov13}. Most of the values in the literature are derived either from the morphology of the mean pulse profile within the hollow-cone emission model~\citep[e.g.][]{rankin93b,gould94} or from the position-angle swing of the
linear polarization across the pulse (the rotating-vector
model)~\citep{lyne88, jk19, FAST_polarization23, Meerkat_RVM}. In both cases,
the observed properties of the pulsar emission depend on the viewing geometry and hence on $\alpha$. The measurement of the obliquity is an inherently model-dependent
procedure, and individual values carry large uncertainties
\citep{MillerHamilton93}. Nevertheless, the results
of different methods and authors turn out to be mutually consistent,
at least in the sense of the resulting distributions of $\alpha$ over
the whole pulsar population \citep{tm98, malov13}.

In theory, angle $\alpha$ does not remain constant during a pulsar's lifetime. The model mentioned above implies:
\begin{equation}
    P\dot P \propto \mu^2(1 + \sin^2\alpha),
\end{equation}
and predicts a secular \emph{decrease} of the obliquity ($\dot\alpha < 0$), i.e. a tendency of the magnetic and rotation axes to align. The characteristic timescale of this alignment coincides with the spin-down time itself,
\begin{equation}
    \tau_\mathrm{align} \simeq 2.2\times 10^{5}\, I_{45}\, P_{0.1}^{2}\, \mu_{30}^{-2}\ \mathrm{yr},
    \label{eq:tausd}
\end{equation}
where $I_{45} = I/(10^{45}\,\mathrm{g\,cm^2})$ is the stellar moment
of inertia, $P_{0.1} = P/(0.1\,\mathrm{s})$ and $\mu_{30} = \mu/(10^{30}\,\mathrm{G\,cm^3})$. Thus, over a few
million years, a typical pulsar is expected to gradually evolve towards
an aligned rotator ($\alpha \to 0$).

The initial distributions of $P_0=P(t=0)$ and $\alpha_0$ are crucial for understanding pulsar population evolution and the statistical properties of the observed ensemble. As for the starting distribution $p(P_0)$ of spin periods, many attempts have sought to reconstruct it. Studies have found consistency with either a log-normal~\citep{Igoshev22} or normal shape~\citep{fgk06, pt12, gullon14, bb25}. In \citet{Igoshev22} and \citet{pt12}, the authors analyzed young pulsars associated with the observed supernova remnants (SNRs), while \citet{fgk06}, \citet{gullon14}, and \citet{bb25} adopted a population synthesis approach. The latter two works also accounted for magnetic obliquity evolution, but assumed a simple isotropic initial distribution $p(\alpha_0) \propto \sin\alpha_0$. In general, the two approaches disagree: the SNR-connected pulsars analysis yields systematically shorter initial periods, $\langle P_0 \rangle \sim 0.1$ s, compared with population synthesis, $\langle P_0 \rangle \sim 0.3$ s. This might be a result of selection effects. For instance, one might speculate that fast proto-NS spin-up during a supernova explosion correlates with a brighter, longer-lived remnant.

In contrast to $P_0$, initial magnetic angles have received much less direct attention. In population studies, $\alpha_0$ is almost universally drawn from an isotropic distribution \citep[e.g.][]{gullon14, Dirson22, bb25}, or the pulsars are treated as orthogonal rotators \citep{fgk06}. This choice is a convenient null hypothesis rather than an established result.

So far, observational constraints on birth magnetic angles are statistical and indirect. \citet{tm98} analyzed polarization-based estimates of $\alpha$ for several hundred pulsars and, after correcting for the beaming fraction, found that their parent distribution is not flat but concentrated towards small angles. They also found that the magnetic angle decreases with characteristic age and interpreted this as alignment on a timescale of $\sim 10^7$~yr. \citet{zjm03} modeled the observed distributions of periods, pulse widths, and magnetic angles with Monte Carlo simulations and favored a parent distribution of $\alpha$ concentrated toward small values over a flat one.
\citet{wj08} did not use individual values of $\alpha$ at all: they showed that the observed fraction of interpulse pulsars and the pulse width--period relation cannot be reproduced with randomly oriented magnetic axes that do not evolve, whereas a random orientation at birth followed by exponential alignment on a timescale of $\sim 7\times10^7$~yr fits the data. \citet{young10} reached a similar conclusion from the dependence of pulse width on characteristic age for $\sim 900$ pulsars, with an alignment timescale of $\sim 10^6$~yr and randomly oriented axes at birth. Finally, \citet{Rookyard15} fitted the rotating-vector model to young $\gamma$-ray-loud pulsars, which should not have had time to align, and nevertheless found their $\alpha$ skewed toward small values. They could reconcile this result with an isotropic birth distribution only by invoking an $\alpha$-dependent radio-beam geometry. The statistics of interpulse pulsars, in turn, constrain a combination of the birth distribution and the evolution law rather than each of them separately \citep{arz17, Novoselov2020}.

Thus, in most of these studies a random orientation at birth is either assumed or found to be compatible with the data within a particular evolutionary model. In contrast, the only sample of pulsars young enough to preserve their birth magnetic angles does not obviously support it. Moreover, in all the works that trace the evolution of $\alpha$, the pulsar age is represented by the characteristic age $\tau_\mathrm{ch} = P/2\dot P$ or is a free parameter of a population model.

Theory does not single out the isotropic distribution either. The obliquity at birth is expected to carry an imprint of the mechanism that builds the large-scale field, and different mechanisms point in different directions. 

Models of magnetar formation can produce nearly orthogonal dipoles
\citep{ReboulSalze21, ReboulSalze22}, dipolar fields shaped by rotation
\citep{Raynaud20, Masada22, White22}, or poloidal fields with a strong
non-axisymmetric component \citep{Barrere22, Barrere23, Barrere25}.
Only in slowly rotating or non-rotating collapse, where the field is amplified stochastically and nothing couples the dipole to the spin axis \citep[e.g.][]{SykesMuller24}, is a random orientation the natural outcome. Whether normal radio pulsars have a preferred magnetic inclination at birth remains unclear.

None of these studies, however, provides a distribution of $\alpha_0$ over a realistic progenitor population. Moreover, the obliquity may be reprocessed shortly after birth. Viscous damping of the free precession of a magnetically deformed star drives the magnetic axis either towards or away from the spin axis, depending on whether the internal field is predominantly poloidal or toroidal \citep{Jones76, DallOssoPerna17, DallOsso18}; coupled with the electromagnetic torque, this tends to sort young neutron stars into nearly aligned and nearly orthogonal rotators early in their lives \citep{LanderJones18, LanderJones20}. The distribution of the ``initial'' $\alpha$ therefore has to be determined empirically.

To our knowledge, no attempt has been made so far to estimate the initial obliquities of individual pulsars based on ages measured independently of the pulsar timing. This work addresses that gap. Our goal is to obtain the empirical distributions of
$\alpha_0$ and $P_0$
of classical isolated radio pulsars. The underlying idea is to consistently integrate their evolution backward in time, starting from the observed values $(\alpha, P, \dot P)$, taking into account the evolution of the magnetic field and using the kinematic ages of the pulsars. In particular, we ask whether the birth obliquity distribution is isotropic ($\propto \sin\alpha_0$, with the magnetic axis randomly oriented over the sphere), uniform ($\propto \mathrm{const}$), or shows a preferred geometry.

The paper is organized as follows. In Section~\ref{sec:sample} we
describe the sample of objects with known magnetic angles and
independent ages. In Section~\ref{sec:model} we formulate the model,
discuss its consistency with the data and describe the numerical
scheme. The results are
presented in Section~\ref{sec:results} and discussed in
Section~\ref{sec:discussion}, which also contains the tests of the
method on mock data. Section~\ref{sec:concl} presents the conclusions.

\section{Data}
\label{sec:sample}

To estimate a pulsar's initial magnetic angle, we need
its current value $\alpha = \alpha(t)$, and the pulsar age $t$.
We adopted the magnetic angle estimates reported by \citet{rankin93b}
and \citet{lyne88}, which are available for a large number of objects. 
The former are derived from the widths of the core
components of the average pulse profiles. The latter combine the pulse width with the maximum rate of the polarization position-angle swing, assuming an empirical relation between the emission-beam radius and the spin period. 
These two methods
have been shown to yield similar distributions of
$\alpha$ \citep{tm98}, although the individual values may differ
considerably \citep{MillerHamilton93}. However, the pulsars under investigation in the following analysis show mostly
``cone-dominated'' pulse profiles. The estimates of $\alpha$ for such objects are more consistent in those two works.

Restricting our selection to isolated normal radio pulsars, we identified estimates of $\alpha$ for 158 objects in these studies. Only six of these objects have characteristic ages
$\tau_\mathrm{ch} = P/(2\dot P)$ below $10^5$~yr and their magnetic angles can be assumed to be close to the initial values. 
Thus, estimating $\alpha_0$
generally requires the pulsar age $t$ and an assumed model of its magneto-rotational evolution. The initial magnetic angle is
then obtained by tracing the pulsar's evolution backward in time over an interval $t$. 
This procedure also yields the initial values of the other parameters.

\begin{table*}[t!]
 \caption{Parameters of the pulsars in the sample}\label{tab:pulsars}
 {\footnotesize\renewcommand{\arraystretch}{1.15}\begin{tabular*}{\textwidth}{@{\extracolsep{\fill}}rlrrrrrccc}
   \toprule
    N & Name & $\alpha$ & $P$ & $\dot P$ & $B$ & $\log\tau_\mathrm{ch}$ &
     $\log t_\mathrm{kin}$ & $\alpha$ Ref. & Kin.\ age Ref.\\
    & & (deg) & (s) & ($10^{-15}$\,s\,s$^{-1}$) & ($10^{12}$\,G) & (yr) &
     (yr) & & \\
     \hline
    1 & J0108+6608 & 26 & 1.28366 & 13.1 & 4.14 & 6.19 & $5.48^{+0.12}_{-0.18}$ & [1] & [5] \\
    2 & J0139+5814$^\dag$ & 42 & 0.272451 & 10.7 & 1.73 & 5.61 & $6.1^{+0.1}_{-0.1}$ & [1] & [3] \\
    3 & J0358+5413 & 51 & 0.156382 & 4.40 & 0.839 & 5.75 & $5.7^{+0.3}_{-0.5}$ & [1] & [3] \\
    4 & J0406+6138 & 83 & 0.594576 & 5.57 & 1.84 & 6.23 & $6.000^{+0.041}_{-0.046}$ & [1] & [5] \\
    5 & J0454+5543 & 24 & 0.340729 & 2.37 & 0.910 & 6.36 & $5.9^{+0.2}_{-0.2}$ & [1] & [3] \\
    6 & J0601-0527 & 51.8 & 0.395969 & 1.30 & 0.727 & 6.68 & $6.54^{+0.18}_{-0.13}$ & [2] & [5] \\
    7 & J0629+2415$^\dag$ & 30 & 0.476623 & 2.00 & 0.987 & 6.58 & $7.04^{+0.15}_{-0.15}$ & [1] & [5] \\
    8 & J0630-2834 & 15.8 & 1.244419 & 7.12 & 3.01 & 6.44 & $6.2^{+0.2}_{-0.1}$ & [2] & [3] \\
    9 & J0659+1414 & 30 & 0.384891 & 55.0 & 4.66 & 5.05 & $5.7^{+0.3}_{-0.4}$ & [1] & [3] \\
    10 & J0729-1836 & 28.2 & 0.51016 & 19.0 & 3.15 & 5.63 & $5.48^{+0.73}_{-0.18}$ & [2] & [5] \\
    11 & J0738-4042 & 17 & 0.37492 & 1.62 & 0.788 & 6.57 & $6.7^{+0.1}_{-0.1}$ & [1] & [3] \\
    12 & J0742-2822 & 37 & 0.166762 & 16.8 & 1.69 & 5.20 & $5.6^{+0.2}_{-0.3}$ & [1] & [3] \\
    13 & J0826+2637 & 84 & 0.530661 & 1.71 & 0.964 & 6.69 & $6.47^{+0.07}_{-0.07}$ & [1] & [4] \\
    14 & J0837+0610 & 60.7 & 1.273768 & 6.80 & 2.98 & 6.47 & $6.1^{+0.5}_{-0.1}$ & [2] & [3] \\
    15 & J0837-4135 & 50 & 0.751624 & 3.54 & 1.65 & 6.53 & $5.3^{+0.2}_{-0.5}$ & [1] & [3] \\
    16 & J0922+0638$^\dag$ & 48 & 0.430627 & 13.7 & 2.46 & 5.70 & $6.2^{+0.3}_{-0.1}$ & [1] & [3] \\
    17 & J0953+0755 & 5.9 & 0.253065 & 0.230 & 0.244 & 7.24 & $6.28^{+0.59}_{-0.16}$ & [2] & [5] \\
    18 & J1136+1551 & 51.3 & 1.187913 & 3.73 & 2.13 & 6.70 & $6.0^{+0.6}_{-0.2}$ & [2] & [3] \\
    19 & J1453-6413 & 56 & 0.179485 & 2.75 & 0.710 & 6.02 & $6.1^{+0.2}_{-0.2}$ & [1] & [3] \\
    20 & J1456-6843 & 37 & 0.263377 & 0.0983 & 0.163 & 7.63 & $7.8^{+0.4}_{-0.4}$ & [1] & [3] \\
    21 & J1509+5531 & 45 & 0.739682 & 5.00 & 1.95 & 6.37 & $6.4^{+0.2}_{-0.4}$ & [1] & [3] \\
    22 & J1559-4438 & 32 & 0.257056 & 1.02 & 0.518 & 6.60 & $6.400^{+0.120}_{-0.097}$ & [1] & [5] \\
    23 & J1645-0317 & 70 & 0.38769 & 1.78 & 0.841 & 6.54 & $6.6^{+0.3}_{-0.1}$ & [1] & [3] \\
    24 & J1709-1640$^\dag$ & 49 & 0.653054 & 6.31 & 2.05 & 6.21 & $6.5^{+0.5}_{-0.1}$ & [1] & [3] \\
    25 & J1740+1311 & 41 & 0.80305 & 1.45 & 1.09 & 6.94 & $6.7^{+0.3}_{-0.1}$ & [1] & [3] \\
    26 & J1741-0840 & 28.7 & 2.043082 & 2.27 & 2.18 & 7.15 & $7.00^{+0.13}_{-0.11}$ & [2] & [5] \\
    27 & J1820-0427 & 65 & 0.598076 & 6.33 & 1.97 & 6.18 & $6.080^{+0.097}_{-0.079}$ & [1] & [5] \\
    28 & J1840+5640 & 56.6 & 1.652862 & 1.49 & 1.59 & 7.24 & $6.52^{+0.27}_{-0.16}$ & [2] & [5] \\
    29 & J1844+1454 & 63 & 0.375463 & 1.87 & 0.848 & 6.50 & $6.0^{+0.1}_{-0.1}$ & [1] & [3] \\
    30 & J1913+1400 & 52 & 0.521473 & 0.804 & 0.655 & 7.01 & $6.83^{+0.16}_{-0.16}$ & [2] & [5] \\
    31 & J1915+1009 & 64 & 0.404547 & 15.3 & 2.51 & 5.62 & $5.6^{+0.2}_{-0.3}$ & [1] & [3] \\
    32 & J1919+0021 & 81 & 1.27226 & 7.67 & 3.16 & 6.42 & $6.300^{+0.061}_{-0.046}$ & [2] & [5] \\
    33 & J1935+1616 & 72 & 0.358738 & 6.00 & 1.48 & 5.98 & $6.2^{+0.1}_{-0.1}$ & [1] & [3] \\
    34 & J1955+5059 & 44 & 0.518938 & 1.37 & 0.854 & 6.78 & $5.9^{+0.1}_{-0.1}$ & [1] & [3] \\
    35 & J2022+2854 & 72 & 0.343402 & 1.89 & 0.816 & 6.46 & $6.3^{+0.2}_{-0.1}$ & [1] & [3] \\
    36 & J2055+3630 & 34 & 0.221508 & 0.369 & 0.289 & 6.98 & $6.980^{+0.100}_{-0.085}$ & [2] & [5] \\
    37 & J2157+4017 & 22.4 & 1.525266 & 3.43 & 2.32 & 6.85 & $6.6^{+0.2}_{-0.1}$ & [2] & [3] \\
    38 & J2219+4754 & 42 & 0.538469 & 2.77 & 1.23 & 6.49 & $6.2^{+0.1}_{-0.1}$ & [1] & [3] \\
    39 & J2225+6535 & 16 & 0.682542 & 9.66 & 2.60 & 6.05 & $5.90^{+0.21}_{-0.20}$ & [2] & [5]\\
   \bottomrule
    \end{tabular*}}
    \par\smallskip
    \begin{minipage}{\linewidth}
    \footnotesize\raggedright
    {\textit{References:}
     [1] \citet{rankin93b};
     [2] \citet{lyne88};
     [3] \citet{nout13};
     [4] \citet{tetz14};
     [5] \citet{Igoshev19}.}\par
    {$^\dag$The pulsars excluded from the analysis as their kinematic ages,
     magnetic angles and timing parameters are inconsistent with the adopted
     spin-down evolution model (Section~\ref{sec:compatibility}).}
    \end{minipage}
\end{table*}

\begin{figure}
\centering
\paperfigure{\columnwidth}{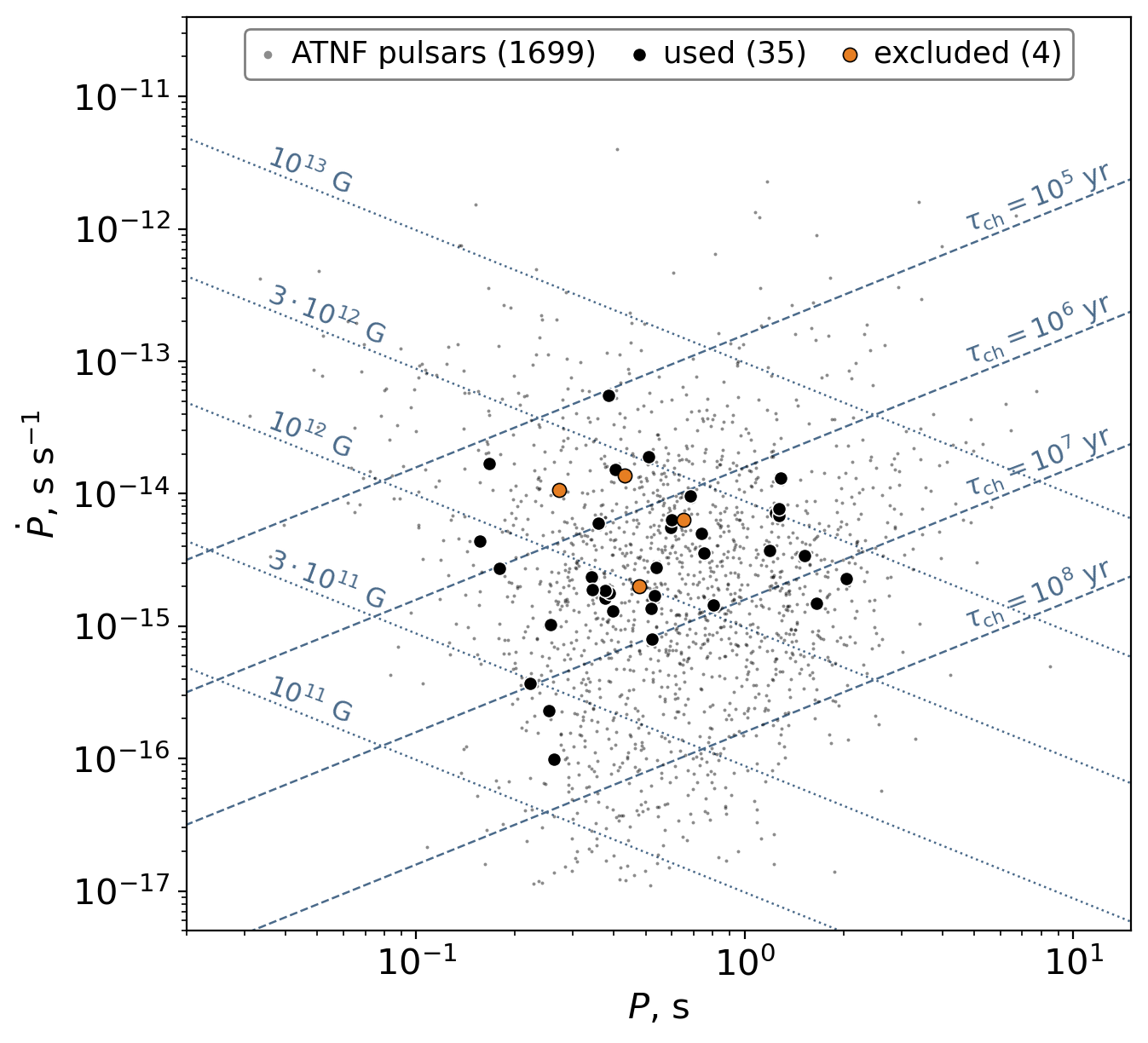}
\caption{The $P$--$\dot P$ diagram. Small dots: isolated normal
pulsars from the ATNF catalog. Black circles: the 35 pulsars used in the analysis. 
Orange circles: the four pulsars whose parameters are inconsistent with the adopted model
(see Section~\ref{sec:compatibility}). The dashed and dotted lines
show constant characteristic age $\tau_\mathrm{ch} = P/2\dot P$ and
constant $B = 3.2\times10^{19}\sqrt{P\dot P}$~G, respectively.}
\label{fig:ppdot}
\end{figure}

Several recent studies have provided pulsar age estimates that
do not rely on the spin period or its derivative, and hence are
independent of the characteristic age $\tau_\mathrm{ch}$. These
include ages found from associated supernova remnants and
so-called kinematic ages. A kinematic age is the time taken by
a pulsar to travel in the Galactic gravitational potential,
from its birthplace in the Galactic plane to its present
position. Age estimates based on supernova remnants are relevant
only to relatively young pulsars and are therefore not used in
our analysis. We compiled kinematic age estimates for 81 pulsars
from \citet{nout13}, \citet{tetz14} and \citet{Igoshev19}.
Of these, 39 also appear in the sample of pulsars with known magnetic
inclination angles. Their parameters are listed in
Table~\ref{tab:pulsars}, and Figure~\ref{fig:ppdot} shows their
positions on the period--period derivative ($P$--$\dot P$) diagram,
together with other known isolated radio pulsars. The sample covers
the bulk of the normal pulsar population, with periods
$0.16$--$2$~s, magnetic fields $B \sim 10^{11}$--$5\times10^{12}$~G and
characteristic ages $10^5$--$4\times10^7$~yr.

During the
analysis, we found that the parameters of four pulsars in this
sample (marked in Table~\ref{tab:pulsars} and Figure~\ref{fig:ppdot})
were inconsistent with the magneto-rotational evolution
model adopted in this work: their kinematic ages are too long for
their spin-down parameters. These objects were excluded from
further consideration; details are given in
Section~\ref{sec:compatibility}. Our final sample therefore
comprises 35 isolated normal radio pulsars.

\section{The model}
\label{sec:model}

As will become clear from the analysis below, pulsar spin
evolution depends on $\alpha$ through $\sin^2\alpha$.
For brevity, we therefore introduce the notation
\begin{equation}
    x(t) \equiv \sin^2\alpha(t).
\end{equation}
Accordingly, $x_0 = \sin^2\alpha_0$, where $\alpha_0$ is the
initial magnetic angle. We also define
\begin{equation}
    b(t) = 2\pi B(t)\sqrt{K},
\end{equation}
which is proportional to the equatorial magnetic field strength
$B(t)$ at time $t$. Here, $K = R^6/(Ic^3)$ depends on the stellar
radius $R$ and moment of inertia $I$. The numerical value of $K$
is not required for our calculations.

\subsection{Magneto-rotational evolution of pulsars}
\label{subsec:magrot}

The spin-down and alignment components of the torque govern
$\dot P$ and $\dot\alpha$, and hence $\dot x$, respectively.
Using the notation
introduced above, we obtain
\begin{equation}
    \begin{cases}
        P\dot P = b^2(1+x), \\
        P^2\dot x = -2b^2x(1-x).
    \end{cases}
    \label{eq:rotation_evolution}
\end{equation}
This system yields an equation for $x_0$:
\begin{equation}
    \dfrac{1}{x_0} + \ln x_0
    = \dfrac{1}{x} + \ln x
    - C(P,x)\int_0^t b^2(t')\,\mathrm{d}t',
    \label{eq:main_eq}
\end{equation}
where
\begin{equation}
    C(P,x) = \dfrac{2(1-x)^2}{P^2x} = \mathrm{const}
    \label{eq:integral}
\end{equation}
is a first integral of the system~\eqref{eq:rotation_evolution}.
Thus, $C(P,x)=C(P_0,x_0)$ remains constant throughout the pulsar's evolution.

Once $x_0$ has been obtained,
Equation~\eqref{eq:integral} yields the corresponding initial
spin period:
\begin{equation}
    P_0 = P \dfrac{1-x_0}{1-x}\sqrt{\dfrac{x}{x_0}}.
    \label{eq:initial_period}
\end{equation}

Given that the oldest pulsars in our sample are $\sim 10^7$~yr old,
the assumption of a constant magnetic field, $b = \mathrm{const}$,
may seem questionable. However, the relatively weak magnetic
fields of normal radio pulsars are generally expected to evolve
only modestly over timescales of $\sim 10^7$--$10^8$~yr
\citep{ip21}.
Qualitatively, magnetic field evolution can be divided into
two stages. The first involves slow decay up to an age of
$t_\mathrm{m} \sim 10^6$~yr, driven by the dissipation of crustal
currents due to phonon scattering. This is followed by rapid
decay, approximately described by a power law and governed
by the impurity content of the crustal lattice
\citep{ip15}.
We therefore describe magnetic field evolution using a simple,
purely phenomenological prescription:
\begin{equation}
    b(t) = b_0\left(1+\dfrac{t}{t_\mathrm{m}}\right)^{-\beta},
    \label{eq:field_evolution}
\end{equation}
where $b_0 = 2\pi B_0\sqrt{K}$ is proportional to the initial
magnetic field strength. Within the adopted model, its value
can be inferred from the observed parameters as
\begin{equation}
    b_0^2 = \frac{P\dot{P}}{1+x}
    \left(1+\frac{t}{t_m}\right)^{2\beta}.
    \label{eq:b0}
\end{equation}
provided that $t_\mathrm{m}$ and $\beta$ are specified.

With Equations~\eqref{eq:field_evolution} and \eqref{eq:b0}, the
integral in Equation~\eqref{eq:main_eq} is
$\int_0^t b^2(t')\mathrm{d}t' = P\dot P\,(1+x)^{-1}\,T$, where
\begin{equation}
    T = \int_0^t \left[\frac{b(t')}{b(t)}\right]^2 \mathrm{d}t'
      = t_{\mathrm{m}}\,\frac{u - u^{2\beta}}{1 - 2\beta},
    \qquad u = 1 + \frac{t}{t_{\mathrm{m}}},
    \quad \beta \neq \tfrac{1}{2},
    \label{eq:teff}
\end{equation}
is an effective age. For $\beta = \tfrac{1}{2}$, the integral gives $T = t_{\mathrm{m}}\,u\ln u$. For a constant field, $T = t$, while for a decaying field, $T > t$. Equation~\eqref{eq:main_eq} then takes
the form
\begin{equation}
    \dfrac{1}{x_0} + \ln x_0
    = \dfrac{1}{x} + \ln x
    - \theta\,\dfrac{(1-x)^2}{x(1+x)},
    \qquad \theta \equiv \dfrac{T}{\tau_\mathrm{ch}}.
    \label{eq:main_eq_theta}
\end{equation}
Hence the initial obliquity depends on the observed quantities only
through the present obliquity and the ratio $\theta$ of the effective
age to the characteristic age: $\alpha_0 = \alpha_0(\alpha, \theta)$.

\subsection{Model-to-data consistency}
\label{sec:compatibility}

The adopted model restricts the possible pulsar ages. In
system~\eqref{eq:rotation_evolution} neither $b$ nor $x$ increases
with time, so 
$P^2 - P_0^2 = 2\int_0^t b^2(1+x)\,\mathrm{d}t' \ge 2P\dot P\,t$,
and therefore
\begin{equation}
    t \le \tau_\mathrm{ch}\left(1 - \dfrac{P_0^2}{P^2}\right)
      < \tau_\mathrm{ch}.
    \label{eq:agelimit}
\end{equation}
The true age of a pulsar cannot exceed its characteristic age,
whatever the parameters of the field decay are. The left-hand side of Equation~\eqref{eq:main_eq_theta},
$1/x_0 + \ln x_0$, reaches its minimum value of unity at $x_0 = 1$ over the entire domain $x_0 > 0$.
Thus, given that $x < 1$, a physical solution exists only if
\begin{equation}
    \theta \le \theta_\mathrm{max}(x) =
    \left(\dfrac{1}{x} + \ln x - 1\right)\dfrac{x(1+x)}{(1-x)^2}.
    \label{eq:thetamax}
\end{equation}
The function $\theta_\mathrm{max}(x)$ varies between $0.896$ (at
$\alpha \approx 28^\circ$) and $1$ (for $\alpha \to 0$ or
$90^\circ$). The limiting case $\theta = \theta_\mathrm{max}$
corresponds to $x_0 = 1$ and $P_0 = 0$, i.e. to an infinitely fast
orthogonal rotator at birth. Since $T \ge t$, the constraint becomes
tighter for a faster field decay.

\begin{figure}[ht]
\centering
\paperfigure{\columnwidth}{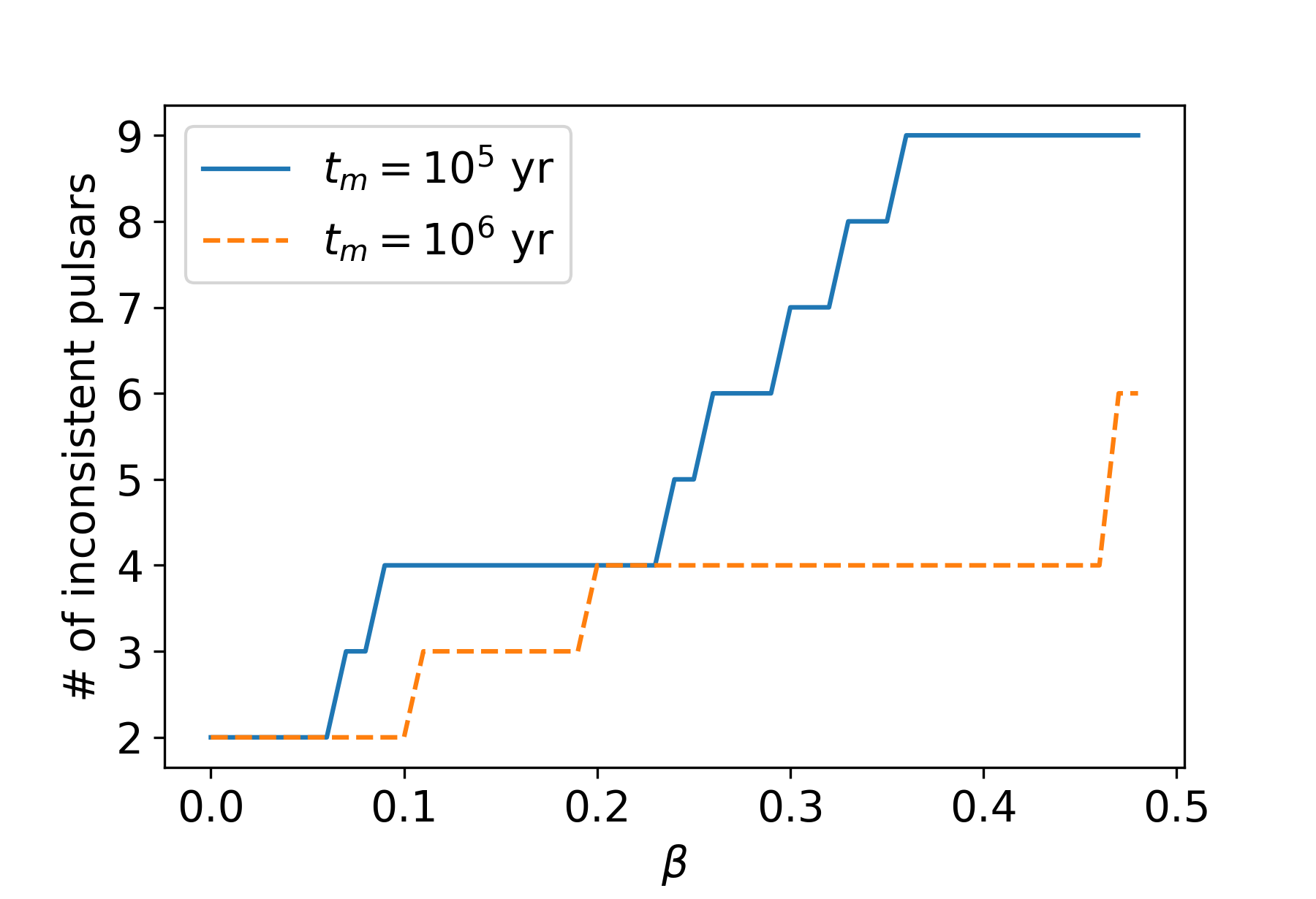}
\caption{Number of pulsars in the sample for which the probability of obtaining a physical solution for $\alpha_0$ is $10^{-4}$ or less. Two curves are shown for $t_\mathrm{m} = 10^5$ and $10^6$~yr across the range $\beta = 0$--$0.5$. Assuming that the magnetic field decay is not extremely fast ($\beta \le 0.2$), we have excluded 4 pulsars from the analysis.}
\label{fig:moc}
\end{figure}

An independent age estimate is not necessarily consistent with this
constraint. For about one third of the 39 pulsars in Table~\ref{tab:pulsars},  the nominal kinematic age exceeds the characteristic age. At these nominal ages, the right-hand side of Equation~\eqref{eq:main_eq_theta} falls below unity, so the equation has no positive real solution for \(x_0\). Therefore, assuming all other parameters fixed, one can estimate for each object the probability of
obtaining a physical solution for a given distribution of $t$. Based on
this, we introduce a \textit{measure of consistency} (m.o.c.) as
\begin{equation}
    \mathrm{m.o.c.}
    = \Pr\!\left[
        \theta(t) < \theta_{\max}
        \,\middle|\,
        P, \dot{P}, x, \beta, t_m
    \right].    
\end{equation}
and consider the observed parameters of a pulsar to be inconsistent
with the model if
\begin{equation}
    \mathrm{m.o.c.} \le 10^{-4}.
\end{equation}
Two of the 39 pulsars, J0139+5814 and J0922+0638, were inconsistent with the model regardless of the magnetic-field evolution parameters~\eqref{eq:field_evolution}; for them, $\mathrm{m.o.c.} < 10^{-5}$. Another two pulsars, J1709$-$1640 and J0629+2415, are consistent with the model for a constant field, but not for $\beta = 0.2$ and $t_\mathrm{m} = 10^5$~yr -- the rapidly decaying field, which we consider below as a limiting setup. Generally, the faster the assumed field decay, the more pulsars in the sample show formal inconsistency with the model (Figure~\ref{fig:moc}). If we assume the decay is not extremely fast ($\beta \le 0.2$), then only these 4 of the 39 pulsars must be excluded from the analysis. Figure~\ref{fig:ppdot} shows them as orange circles. Thus, our final sample consists of 35 objects.
The consistency analysis can also be read as a constraint on the magnetic field evolution of isolated pulsars. As long as their spin-down is described by the model adopted in this work, $\beta \le 0.2$ and $t_\mathrm{m} \ge 10^5$~yr may be taken as a conservative estimate of the parameters of $b(t)$. We note, however, that several of the retained pulsars are consistent with the model only in the lower tail of their age distributions (e.g. J0659+1414 and J0742$-$2822, whose nominal kinematic ages exceed $\tau_\mathrm{ch}$ by factors of $4.5$ and $2.5$). This indicates that either some kinematic ages are overestimated or these pulsars' spin-down deviates from system~\eqref{eq:rotation_evolution}, e.g., due to an effective braking index below three or dipole field growth.

\subsection{Calculations}
\label{sec:calculations}

The initial parameters were calculated by a Monte Carlo procedure
that propagates the uncertainties of the two least certain
quantities, the age and the present obliquity. For each of the 35
pulsars we generated $N = 10^5$ trials. In each trial:

(i) The age was drawn from a split-normal distribution in
$\log t_\mathrm{kin}$, with the mean and the upper and lower standard
deviations listed in Table~\ref{tab:pulsars}. In accordance with
Equation~\eqref{eq:agelimit}, the distribution was truncated from above
at $\log\tau_\mathrm{ch}$.

(ii) The present obliquity was drawn from a uniform distribution within
$\pm 10^\circ$ of the published value; values falling outside the
interval $[0^\circ, 90^\circ]$ were reflected back into it. Neither
\citet{rankin93b} nor \citet{lyne88} provide uncertainties of the individual
values of $\alpha$. Both methods rely on an assumed universal relation
between the size of the emission beam (or of its core) and the period,
so that the error of $\alpha$ grows with $\alpha$ itself, roughly as
$\sigma_\alpha \sim \tan\alpha\,(\sigma_W/W)$, where $W$ is the measured
width; the two sets of estimates agree with each other for
$\alpha \lesssim 40^\circ$ and are essentially uncorrelated at larger
angles \citep{MillerHamilton93}. The adopted smearing is a
representative, and probably optimistic, measure of these errors. It is
comparable to half of the bin width used below.

(iii) Equation~\eqref{eq:main_eq_theta} was solved for $x_0$. In terms
of $y = 1/x_0$ it reads $y - \ln y = G$, where $G$ is its right-hand
side, and its solution is $y = -W_{-1}(-e^{-G})$, where $W_{-1}$ is
the lower branch of the Lambert function. If the drawn pair
$(t,\alpha)$ violated condition~\eqref{eq:thetamax}, a new pair was
drawn. For $\beta = 0$, more than 75 percent of the first draws are
accepted for all pulsars but two (J0738$-$4042 and J2055+3630, about
50 and 60 percent). For $\beta = 0.2$ the acceptance is lower and falls
to $\sim 1$ per cent for J0738$-$4042, J1935+1616 and J2055+3630.

(iv) The initial period was obtained from
Equation~\eqref{eq:initial_period}.

Each trial thus provides a set of 35 values of $\alpha_0$ and $P_0$,
i.e., one realization of the initial distributions. The calculations
were performed for two versions of the field evolution: a constant
magnetic field ($\beta = 0$) and a relatively rapidly decaying field
($\beta = 0.2$, $t_\mathrm{m} = 10^5$~yr). In the latter case, the
magnetic field strength decreases to approximately 40 percent of its
initial value over $10^7$~yr.

The observed sample is affected by a geometric selection: the
probability that the emission beam of a pulsar sweeps across the
observer depends on $\alpha$ and $P$. For two symmetric conal beams of
angular radius $\rho$, this probability (the beaming fraction) is
\citep{tm98}
\begin{equation}
    f(\alpha, P) =
    \begin{cases}
        2\sin\alpha\sin\rho, & \rho < \alpha < 90^\circ - \rho,\\
        \cos(\alpha-\rho), & \alpha > \rho,\ \alpha > 90^\circ - \rho,\\
        1 - \cos(\alpha+\rho), & \alpha < \rho,\ \alpha < 90^\circ - \rho,\\
        1, & 90^\circ - \rho < \alpha < \rho,
    \end{cases}
    \label{eq:beaming}
\end{equation}
where we adopt $\rho = 5.75^\circ P^{-1/2}$ with $P$ in seconds. To
correct for this selection, each pulsar enters all the histograms
with the weight $w = 1/f$ calculated for its observed $\alpha$ and
$P$. 

In each trial we built the weighted histograms of $\alpha_0$, with
four bins of $22.5^\circ$, and of $\log P_0$, with bins of $0.5$~dex,
both normalized to the total number of 35 pulsars. The spread of the
number of pulsars in a bin over the trials is shown below by a
``box plot'', representing its 25 and 75\% percentiles, the median value, and the 
minimal and maximal values over the trials. To compare the
distribution of $\alpha_0$ with a model one, we calculated for each bin
the fraction of trials in which the number of pulsars in this bin,
$N_\mathrm{MC}$, is not less than the model number:
$p_\mathrm{uni} = p(N_\mathrm{MC} \ge N_\mathrm{uni})$ for the
distribution uniform in angle and
$p_\mathrm{iso} = p(N_\mathrm{MC} \ge N_\mathrm{iso})$ for the
isotropic one. A small value of $p$ indicates a lack of pulsars in the bin and a
value close to unity indicates an excess. These quantities
describe the uncertainty of the reconstruction caused by the errors of
$t$ and $\alpha$ only. They do not include the statistical noise of a
sample of 35 objects and are not formal significance levels (see
Appendix below). For the initial periods, we calculated in
each trial the weighted mean and standard deviation of the 35 values of
$P_0$, $(m_{P_0}, s_{P_0})$, and of $\log P_0$,
$(m_{\log P_0}, s_{\log P_0})$, and considered the distributions of
these moments over the trials.

\section{Results}
\label{sec:results}

\subsection{Initial pulsar magnetic inclinations}
\label{sec:res_alpha}

\begin{figure*}[!t]
\centering
\paperfigure{0.48\textwidth}{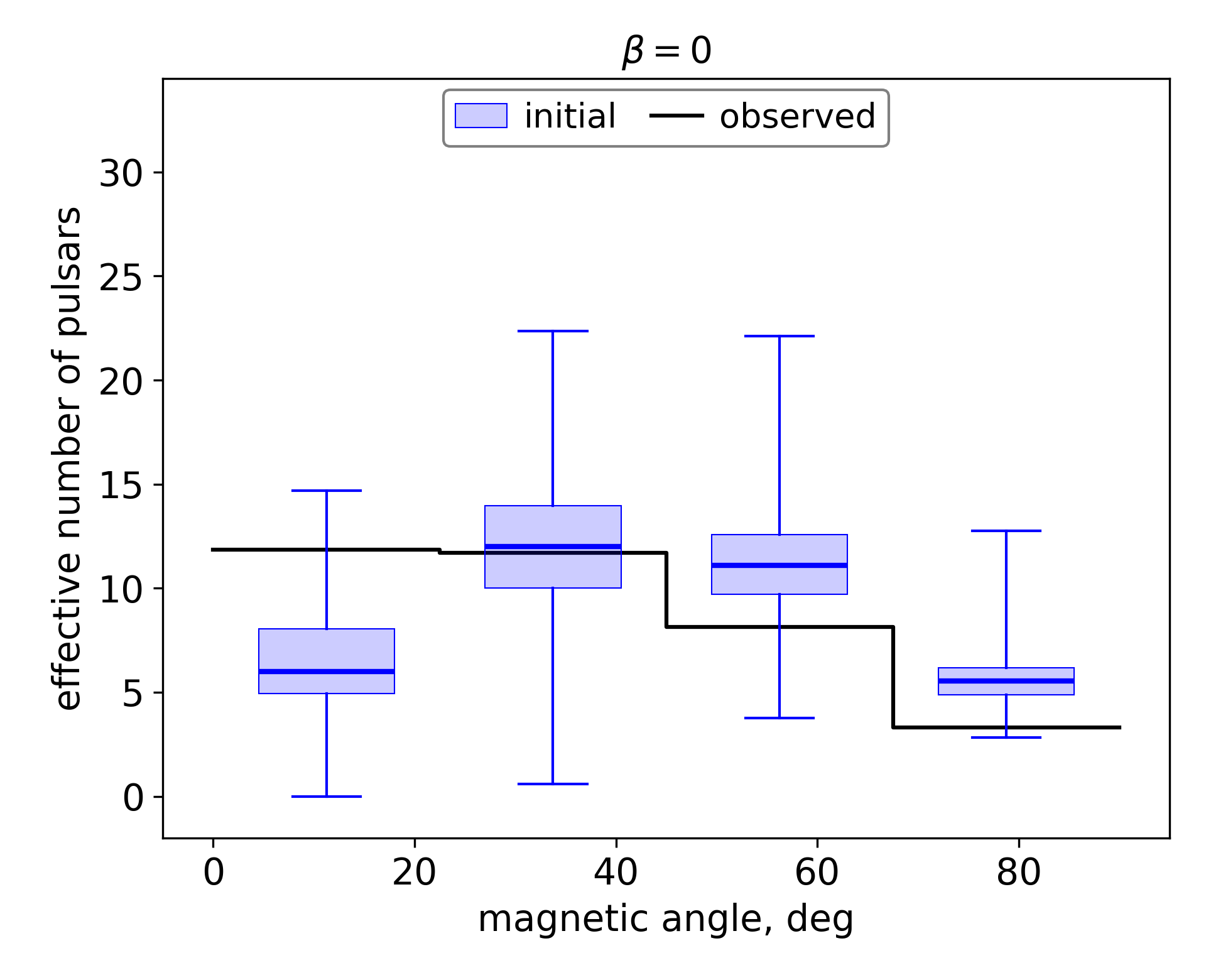}
\paperfigure{0.48\textwidth}{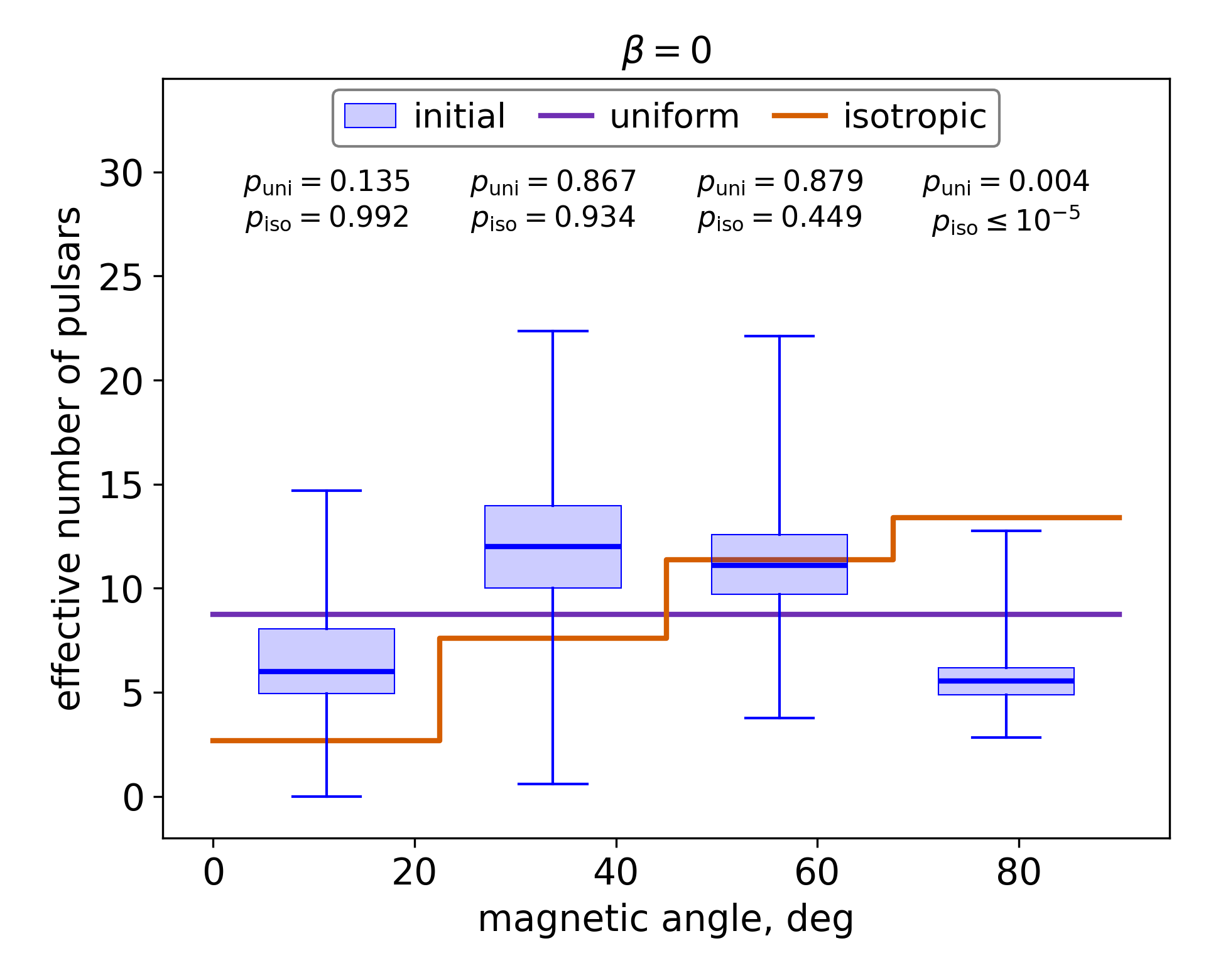}\\[0.1em]
\paperfigure{0.48\textwidth}{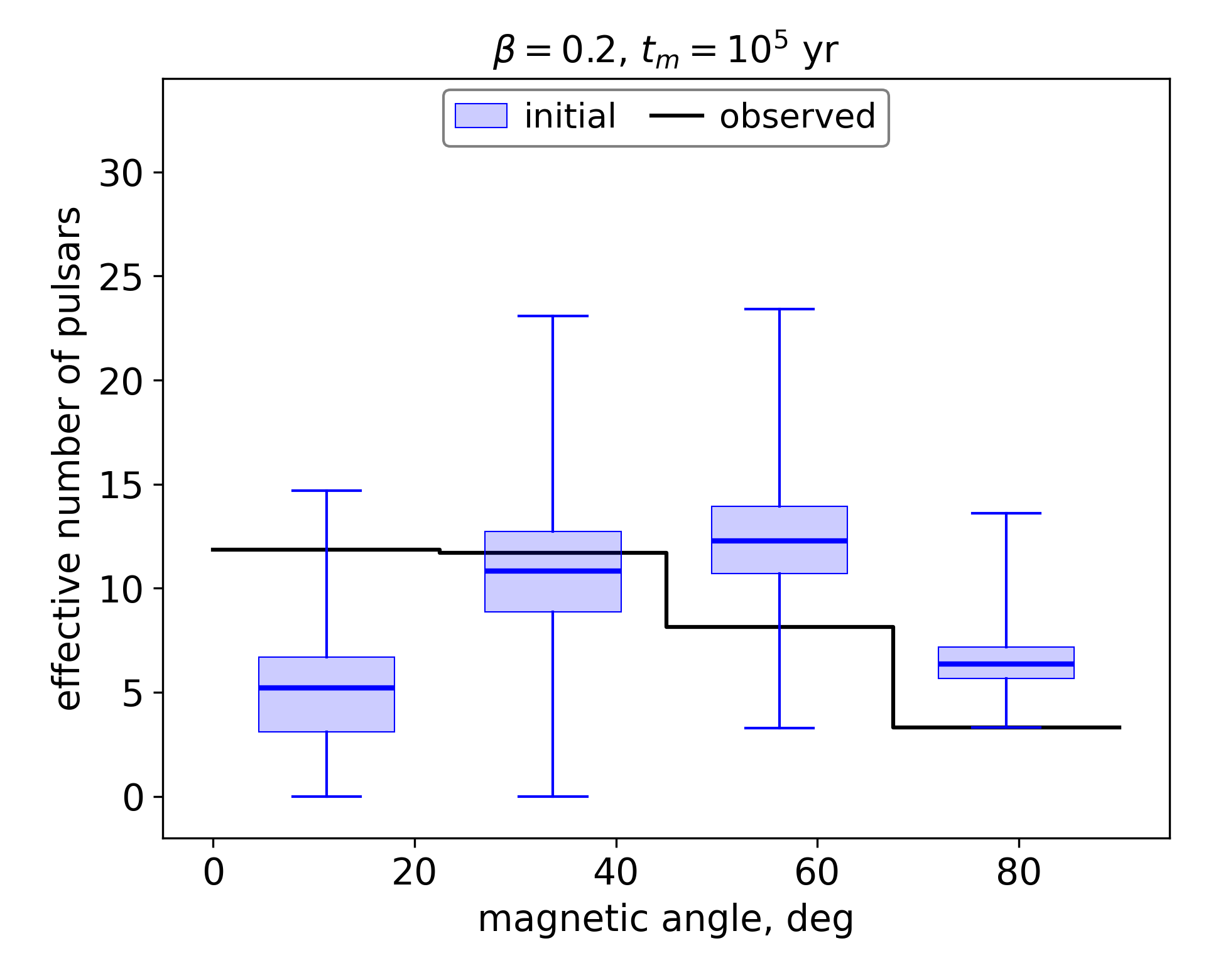}
\paperfigure{0.48\textwidth}{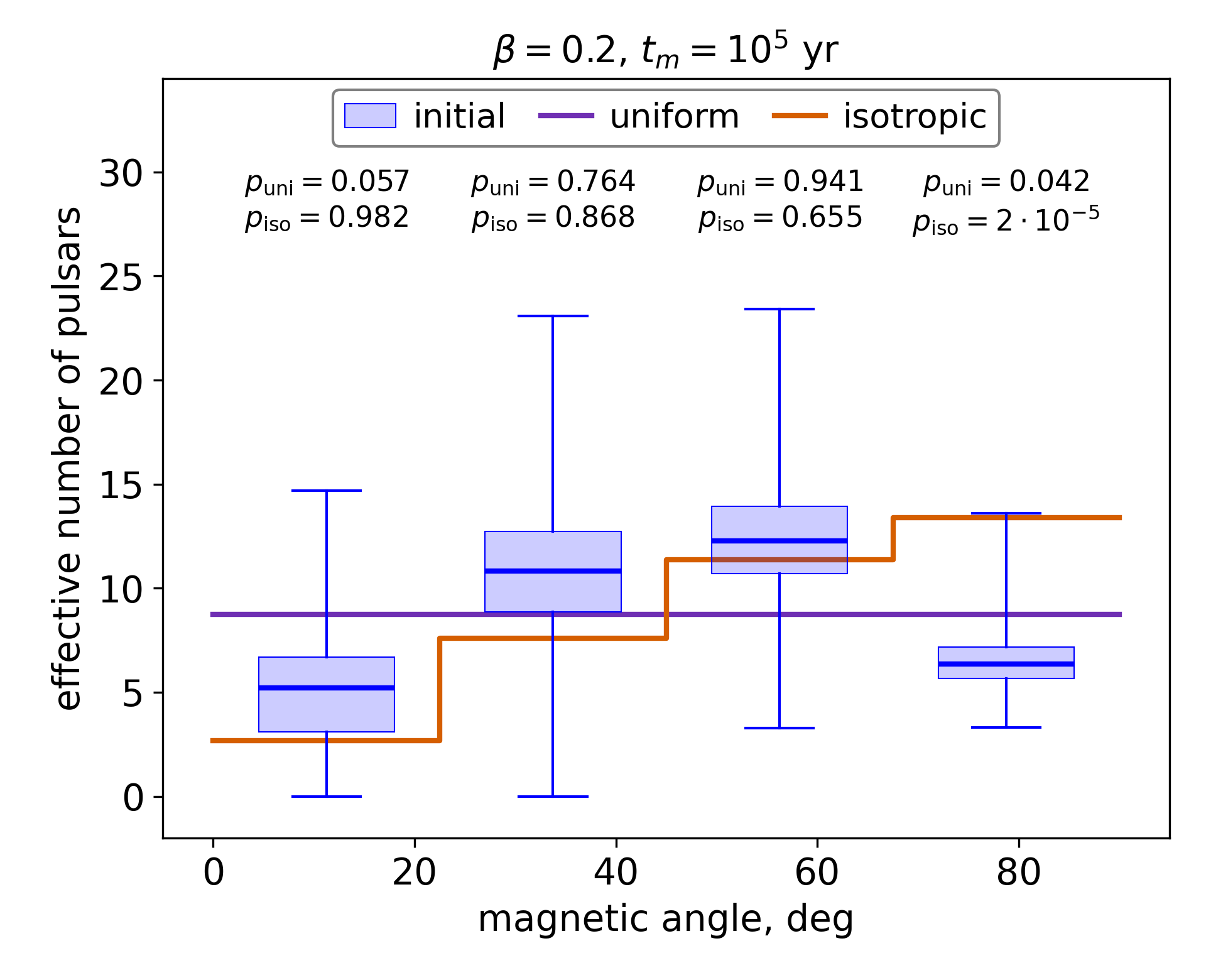}
\caption{
Distributions of the magnetic angles of the 35 pulsars of the final
sample, corrected for the beaming fraction. The upper row assumes a
constant magnetic field ($\beta=0$); the lower row assumes magnetic
field decay with $\beta=0.2$ and $t_\mathrm{m}=10^5$~yr.
The boxes show the distributions of the number of pulsars with the
initial angle $\alpha_0$ in a given bin over the Monte Carlo trials,
which reflect the uncertainties of the kinematic ages and of the
observed $\alpha$; the horizontal bars mark the medians.
Left: comparison with the distribution of the present-day angles
$\alpha$ (black line).
Right: comparison with the distribution uniform in angle (purple) and
with the isotropic distribution, $p(\alpha_0)\propto\sin\alpha_0$
(red). The numbers above the bins are the probabilities
$p_\mathrm{uni}$ and $p_\mathrm{iso}$ that the reconstructed number of
pulsars reaches the model value.
}
\label{fig:angles}
\end{figure*}

The reconstructed distributions of $\alpha_0$ are shown in
Figure~\ref{fig:angles}, and the numbers of pulsars per bin are given in
Table~\ref{tab:alpha0}. The left-hand panels of the figure compare them
with the distribution of the present-day values, corrected for the
beaming in the same way. In the adopted model, the magnetic angle can only
decrease with time, so that the reconstruction shifts the distribution
towards larger angles: the number of pulsars with
$\alpha < 22.5^\circ$ drops from $\approx 12$ to $5$--$6$, while the
number of those with $\alpha > 67.5^\circ$ increases from $3.3$ to
$5.5$ ($\beta = 0$) or $6.4$ ($\beta = 0.2$). The shift is moderate
because for many pulsars $t_\mathrm{kin}$ is well below
$\tau_\mathrm{ch}$, and their $\alpha$ have changed little.

\begin{table}
\normalfont\rmfamily\footnotesize
\caption{Numbers of pulsars in the bins of the magnetic angle
(normalized to 35): observed, expected for the uniform and isotropic
distributions, and reconstructed (median and 16--84 percent range over
the trials), with the probabilities
$p = p(N_\mathrm{MC} \ge N_\mathrm{model})$.}
\label{tab:alpha0}
\centering
\setlength{\tabcolsep}{3pt}
\begin{tabular}{@{}lcccc@{}}
\toprule
 & $0$--$22.5^\circ$ & $22.5$--$45^\circ$ & $45$--$67.5^\circ$ & $67.5$--$90^\circ$\\
\hline
Observed $\alpha$ & 11.9 & 11.7 & 8.2 & 3.3\\
Uniform & 8.75 & 8.75 & 8.75 & 8.75\\
Isotropic & 2.66 & 7.59 & 11.35 & 13.39\\
\hline
\multicolumn{5}{l}{$\beta = 0$}\\
\hline
$\alpha_0$ & $6.0^{+2.3}_{-2.9}$ & $12.0^{+2.8}_{-2.9}$ & $11.1^{+2.2}_{-2.0}$ & $5.5^{+1.0}_{-0.8}$\\
$p_\mathrm{uni}$ & 0.14 & 0.87 & 0.88 & 0.004\\
$p_\mathrm{iso}$ & 0.99 & 0.93 & 0.45 & $\le 10^{-5}$\\
\hline
\multicolumn{5}{l}{$\beta = 0.2$, $t_\mathrm{m} = 10^5$~yr}\\
\hline
$\alpha_0$ & $5.2^{+2.6}_{-2.1}$ & $10.8^{+2.8}_{-2.9}$ & $12.3^{+2.5}_{-2.3}$ & $6.4^{+1.2}_{-1.0}$\\
$p_\mathrm{uni}$ & 0.06 & 0.76 & 0.94 & 0.04\\
$p_\mathrm{iso}$ & 0.98 & 0.87 & 0.66 & $2\times10^{-5}$\\
\specialrule{\heavyrulewidth}{0pt}{0pt}
\end{tabular}
\end{table}

For both versions of the field evolution, the distribution of
$\alpha_0$ is concentrated at intermediate angles: about two thirds of
the (weighted) sample start their evolution with
$22.5^\circ < \alpha_0 < 67.5^\circ$. The comparison with the two
reference distributions (right-hand panels of Figure~\ref{fig:angles})
leads to the following conclusions.

In the three bins below $67.5^\circ$, the reconstructed numbers are
compatible with the distribution uniform in angle
($0.06 \le p_\mathrm{uni} \le 0.94$). With respect to the isotropic
distribution, there is an excess of nearly aligned rotators,
$\alpha_0 < 22.5^\circ$ ($5$--$6$ pulsars against the expected $2.7$,
$p_\mathrm{iso} \approx 0.98$--$0.99$). This excess, however, rests on
a few pulsars with small observed $\alpha$ and large weights: J0953+0755
($\alpha = 5.9^\circ$) alone contributes $3.1$ to this bin.

The most pronounced feature is the lack of pulsars with large initial
obliquities. Only $5.5^{+1.0}_{-0.8}$ ($\beta=0$) or
$6.4^{+1.2}_{-1.0}$ ($\beta=0.2$) pulsars have
$\alpha_0 > 67.5^\circ$, against $13.4$ expected for the isotropic
distribution and $8.75$ for the uniform one. The isotropic number is
reached in none of the $10^5$ trials for $\beta = 0$ and in two of them
for $\beta = 0.2$, and the uniform number in $0.4$ and $4$ per cent of
the trials, respectively. Thus neither of the two reference
distributions describes the whole range of angles: the reconstructed
distribution is flatter than the isotropic one and falls off towards
$90^\circ$.

We conclude that there is evidence for a lack of pulsars with large
(close to $90^\circ$) initial magnetic angles: at the start of
their evolution as radio pulsars, neutron stars are rarely nearly
orthogonal rotators. The faster field decay makes the deficit somewhat
smaller but does not remove it. The strength of this evidence, with
allowance for the finite size of the sample and for the properties of
the method, is assessed in the Appendix.

\subsection{Initial spin periods}
\label{sec:res_p0}
The important disclaimer is that we do \textit{not} consider the resultant distribution of $P_0$ to be trustworthy from 
the physical point of view. Instead, as we show in the Appendix, this distribution is strongly
biased toward the longer periods due to the properties of our numerical scheme. And due to large 
uncertainties in the kinematic ages. In contrast, we can rely on the initial magnetic angles.

\begin{figure}[t]
\centering
\paperfigure{1.0\columnwidth}{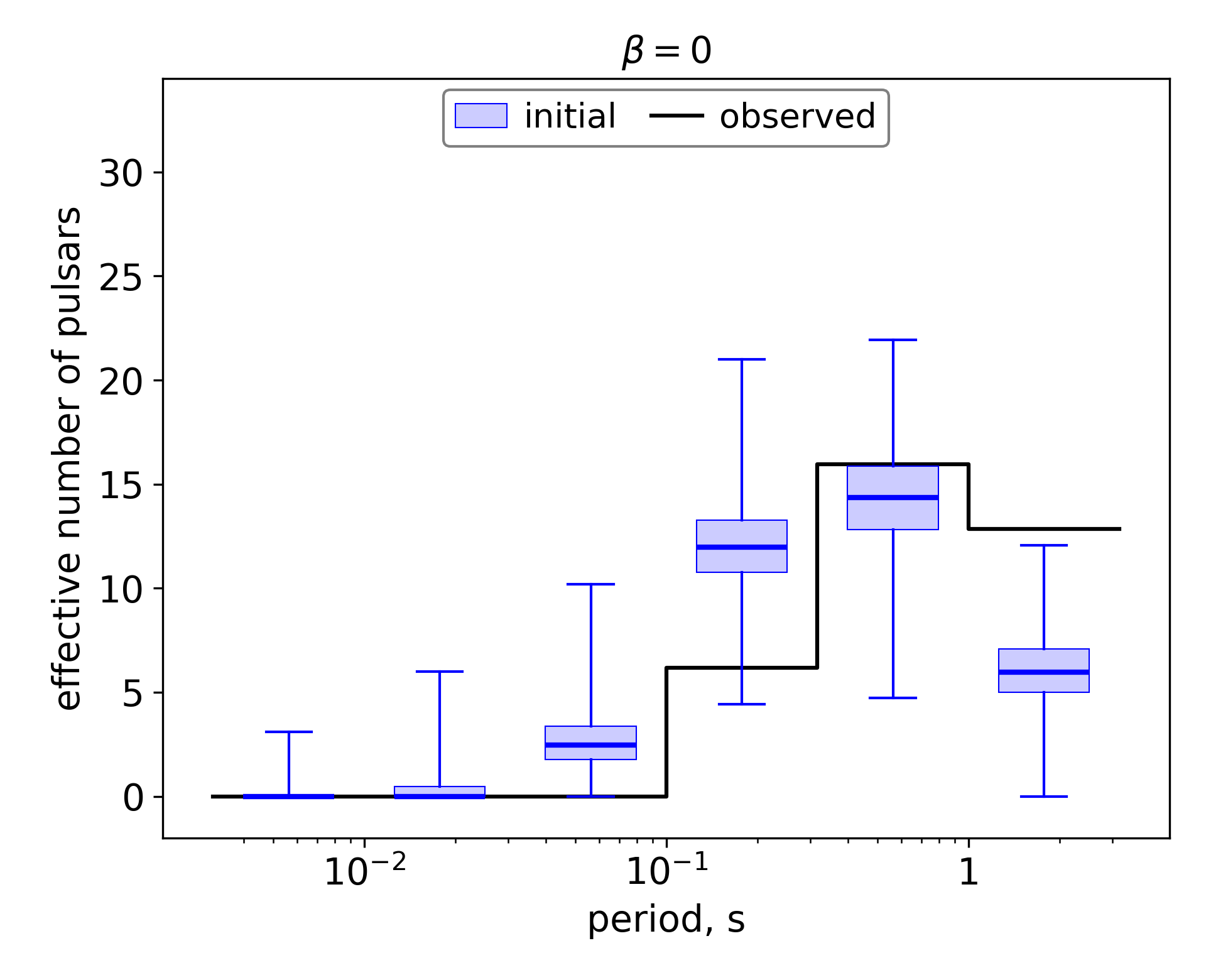}\hfill

\paperfigure{1.0\columnwidth}{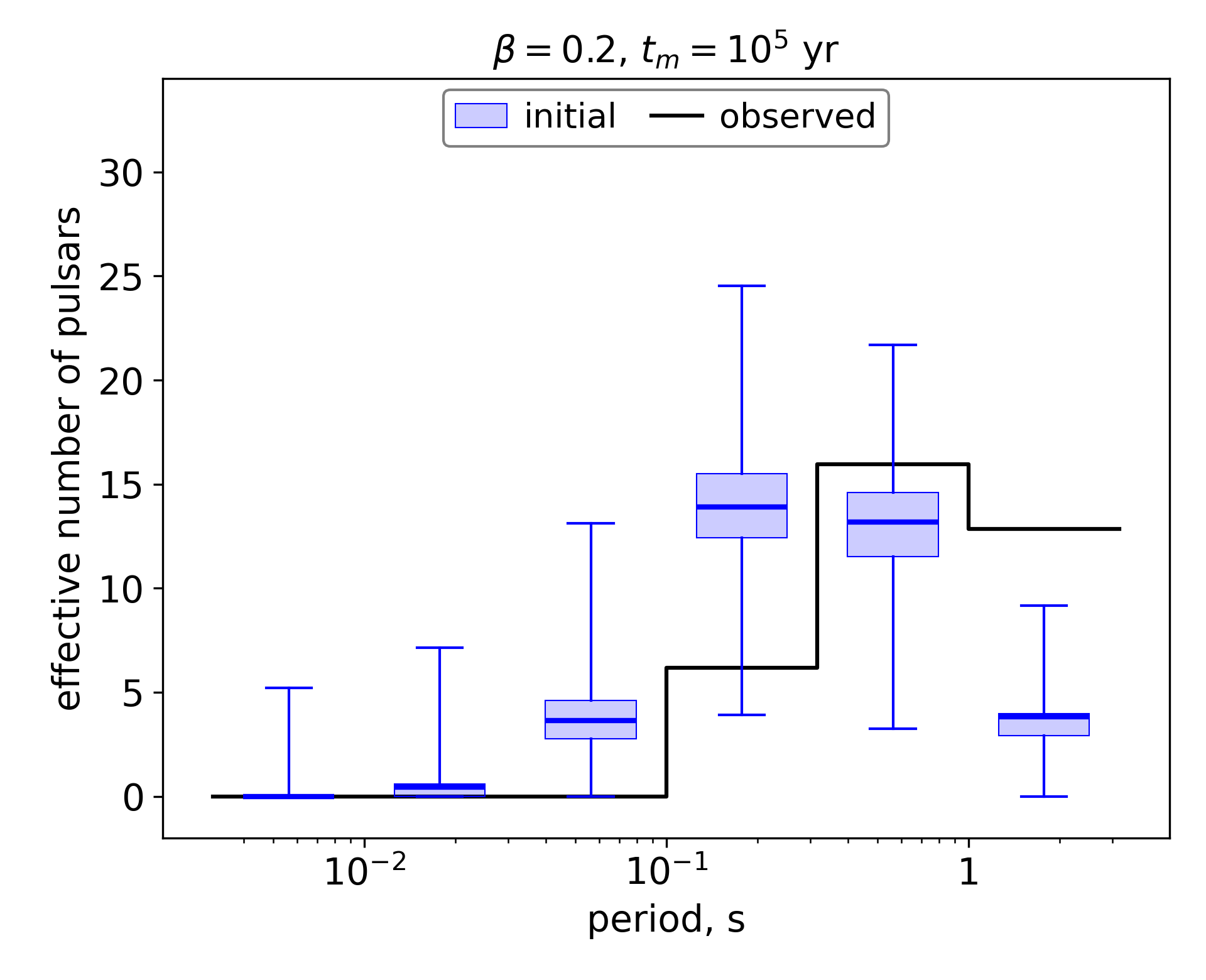}\hfill

\caption{
Initial spin periods of the 35 pulsars, corrected for the beaming
fraction. The upper plot assumes a constant magnetic field ($\beta=0$);
the lower one adopts $\beta=0.2$ and $t_\mathrm{m}=10^5$~yr.}
\label{fig:initial_periods}
\end{figure}

The reconstructed distributions of the initial periods are shown in the
upper row of Figure~\ref{fig:initial_periods} together with the
(weighted) distribution of the present-day periods. As expected, the
initial periods are shorter than the observed ones: the pulsars with
$P > 1$~s, which make up more than a third of the weighted observed
sample, start in about half of the cases from $P_0 < 1$~s. Most of the reconstructed
initial periods fall in the range $0.1$--$1$~s, and only a few pulsars
are compatible with $P_0 < 0.1$~s.

The mean and the standard deviation of the initial periods are
$m_{P_0} = 0.52^{+0.03}_{-0.04}$~s and
$s_{P_0} = 0.39 \pm 0.03$~s for the constant field, and
$m_{P_0} = 0.44 \pm 0.03$~s and $s_{P_0} = 0.35 \pm 0.03$~s for the
decaying one (the median and the 16--84 per cent range over the
trials). For the logarithms of the periods we obtain
$m_{\log P_0} = -0.43^{+0.03}_{-0.03}$,
$s_{\log P_0} = 0.38^{+0.04}_{-0.03}$ and
$m_{\log P_0} = -0.52^{+0.03}_{-0.04}$,
$s_{\log P_0} = 0.40^{+0.04}_{-0.03}$, respectively, i.e. geometric
means of $0.37$ and $0.30$~s.

The
normal distributions of \citet{fgk06} ($0.30 \pm 0.15$~s, FK06 hereafter) and of
\citet{pt12} ($0.10 \pm 0.10$~s), as well as the log-normal distribution
of \citet{Igoshev22} ($m_{\log P_0} = -1.04$, $s_{\log P_0} = 0.53$,
i.e. a geometric mean of $0.09$~s), all lie far outside the regions
occupied by the Monte Carlo trials. Formally, the initial periods
reconstructed here are therefore considerably longer than those
obtained earlier: the mean is larger than that of FK06 by a factor of
$1.4$--$1.7$ and the geometric mean exceeds that of I+22 by a factor of
$3$--$4$, while the dispersion of $\log P_0$ is close to that of I+22. A faster
field decay reduces the reconstructed periods, but not sufficiently to
remove the difference.

\section{Discussion: Are initial magnetic angles primordial or reprocessed?}
\label{sec:discussion}

After correcting for geometric selection, the reconstructed distribution of initial magnetic angles is not isotropic. Relative to $p(\alpha_0) \propto \sin\alpha_0$, the sample contains too few pulsars with $\alpha_0 > 67.5^\circ$ and too many with $\alpha_0 < 45^\circ$ (Figure~\ref{fig:angles}, Table~\ref{tab:alpha0}). The uniform distribution describes the data better, but it has no particular physical meaning and serves only as a reference. Thus, we conclude that nearly orthogonal rotators are under-represented at the start of the evolution. Two remarks are in order. First, in an aligning model, the backward integration can only increase the angle, $\alpha_0 \ge \alpha$. The deficit of large $\alpha_0$ is therefore already present in the observed sample and is merely reduced by the reconstruction: any model with slower alignment, and the counter-aligning scenario in particular, would give even smaller $\alpha_0$ and a stronger departure from isotropy. Only substantially faster alignment than in system~\eqref{eq:rotation_evolution} could restore it. Second, the evidence is moderate (see Appendix below): it rests on 35 objects, a correction for the beaming fraction, and $\alpha$ estimates of limited accuracy. With these reservations, we discuss below what such a deficit would imply.

\subsection{Can the electromagnetic torque have reprocessed the birth obliquity?}

Hypothetically, the lack of strongly oblique rotators could result from a rapid decrease in $\alpha$ in a newborn neutron star under the same electromagnetic torque that operates later. Within the adopted model, this possibility is tightly constrained, because the torque does not change $\alpha$ and $P$ independently.
A star born with $(P_\mathrm{i}, x_\mathrm{i})$ that has reached the obliquity $x_0 < x_\mathrm{i}$ has necessarily spun down to $P_0 > P_\mathrm{i}$
accordingly to the Equation (\ref{eq:initial_period}). For example, alignment from $80^\circ$ to $60^\circ$ requires $P_0 \approx 10\,P_\mathrm{i}$.

However, if the pulsars are born with periods of an order of one tenth of a second, as widely suggested 
then for $\mu_{30} \sim 1$ Equation~\eqref{eq:tausd} gives $\tau_\mathrm{align} \lesssim 10^6$~yr. This is comparable to the ages of most pulsars in the sample, and this slow alignment is exactly what our backward integration already accounts for. A rapid electromagnetic reprocessing of $\alpha$ is then excluded: a neutron star born as a slow rotator keeps its birth obliquity for $\sim 10^6$~yr, and the reconstructed distribution of $\alpha_0$ must be primordial or shaped by a non-electromagnetic process. 

The contradiction disappears only if $P_0$ and $\alpha_0$ are not the birth values but the values at the end of a short early episode of enhanced torque. 
For instance, it could be driven by either a dipole field initially several times stronger than the present one that decays within $\lesssim 10^3$--$10^4$~yr, an extra torque of a different nature (a proto-neutron-star wind or interaction with fallback matter), or a combination of the two. 
The smooth law~\eqref{eq:field_evolution} cannot represent such an episode, and kinematic ages of $\gtrsim 10^5$~yr cannot resolve it, so that our reconstruction would stop at its end. Formally, every curve $C = \mathrm{const}$ passes through $(P = 0, x = 1)$, so each pulsar in the sample is compatible with birth as a nearly orthogonal millisecond rotator, as predicted for the magnetorotational dynamo. However, this would require magnetar-strength fields in all of them and the release of $\gtrsim 10^{50}$~erg of rotational energy into every supernova remnant.

This can, in principle, be distinguished observationally. Pulsars younger than the duration of the enhanced-torque episode should be, on average, both faster and more oblique than the values of $P_0$ and $\alpha_0$ derived here, and their magnetic angles should decrease on a measurable time scale. The existing data do not support this in an obvious way: young $\gamma$-ray-loud pulsars with characteristic ages below $\sim 5\times10^5$~yr already show a preference for small $\alpha$ \citep{Rookyard15}, and the obliquity of the Crab pulsar is currently increasing rather than decreasing \citep{Lyne13}. A homogeneous set of $\alpha$ for pulsars associated with supernova remnants would provide a direct test.

\subsection{Non-electromagnetic reprocessing}

The obliquity can also change through the dissipation of free precession in the stellar interior, which does not require any spin-down. For a star deformed by a dominant toroidal field, this mechanism increases $\alpha$ and was predicted to produce a bimodal distribution with few objects at intermediate angles, $\alpha \sim 20^\circ$--$70^\circ$, for birth periods of $10$--$300$~ms and a wide range of field strengths \citep{DallOssoPerna17}. Our reconstruction shows the opposite: most pulsars start their evolution precisely in this interval, and nearly orthogonal rotators are rare. Efficient viscous orthogonalization of ordinary pulsars is therefore disfavoured, consistent with the expectation that for sub-magnetar fields and moderate rotation, buoyancy forces stall this process and preserve intermediate $\alpha$ \citep{LanderJones18}. If the internal field is instead predominantly poloidal, the same dissipation acts towards alignment and could contribute to the deficit of large angles without affecting the period.

\subsection{Implications of a primordial origin}

If the distribution is primordial, it carries information on the origin of the field. For $P_0 \sim 0.1$--$1$~s, the rotation period exceeds the convective overturn time of a proto-neutron star by two to three orders of magnitude, so that rotation cannot organize a dynamo. The stochastic field amplification seen in simulations of non-rotating collapse \citep{SykesMuller24} should result in a random, i.e., isotropic, orientation of the dipole. A lack of orthogonal rotators among slowly rotating neutron stars is therefore not trivial. It suggests that the dipole axis retains a memory of the rotation axis: either the field is inherited from the progenitor core, where it was organized by rotation, or it was generated while the proto-neutron star rotated much faster than $P_0$, with the subsequent spin-down caused by a torque that does not change $\alpha$. It also disfavors the picture in which the equatorial dipole produced by the magnetorotational instability \citep{ReboulSalze21, ReboulSalze22} is the dominant outcome for ordinary pulsars. If the spin of the neutron star is set by late fallback \citep{Janka22}, independently of the field already frozen into the star, the orientation would again be random; a non-isotropic distribution then requires the field to be generated by the same fallback-induced shear, as in the Tayler--Spruit scenario \citep{Barrere22}.

A caveat applies to all the above. Recent three-dimensional magneto-thermal simulations indicate that the dipole axis of a neutron star with a strong small-scale crustal field can drift by tens of degrees within $\sim 10^3$~yr \citep{DehmanVigano26}. If such a drift is common, obliquity is not an integral of early evolution; the quantity reconstructed here is then an effective starting value for secular magnetospheric evolution, and any primordial anisotropy is partially erased.

\section{Conclusions}
\label{sec:concl}

We have estimated the initial magnetic angles and spin periods of
35 isolated radio pulsars with known kinematic ages by integrating
their magneto-rotational evolution backward in time. The main results
are as follows.

(i) The adopted spin-down model requires the true age of a pulsar to be
smaller than its characteristic age. The nominal kinematic ages of
about one third of the pulsars with known $\alpha$ violate this
condition, and four pulsars had to be excluded from the analysis. The
consistency of the remaining ones with the model limits the field decay
to $\beta \le 0.2$ for $t_\mathrm{m} \ge 10^5$~yr.

(ii) The distribution of the initial angles, corrected for the
beaming fraction, is concentrated at intermediate angles. There is a
lack of pulsars with $\alpha_0 > 67.5^\circ$: $5.5$--$6.4$ of 35
against $13.4$ expected for the isotropic distribution and $8.75$ for
the distribution uniform in angle.

(iii) Tests on mock samples show that individual initial $\alpha_0$
are recovered without a significant bias, while the method itself
removes on average $3$--$4$ pulsars from the last bin. The observed
deficit is stronger than that, but an isotropic parent distribution
produces a similar one in $\sim 10$ per cent of the cases. The lack of
nearly orthogonal rotators at birth is thus indicated but not firmly
established.

(iv) If real, this deficit is not expected for slowly rotating
proto-neutron stars with stochastically amplified fields. Within the
adopted model it cannot result from a rapid electromagnetic alignment,
unless both $\alpha_0$ and $P_0$ refer to the end of a short early
episode of enhanced torque.

(v) The reconstructed initial periods ($m_{P_0} = 0.4$--$0.5$~s,
$m_{\log P_0} = -0.4$ to $-0.5$) are formally longer than those obtained
in earlier works, but the method overestimates $P_0$ for the pulsars
that have spun down considerably. The data are not in conflict with the
distribution of \citet{fgk06}.

The isotropic distribution of the initial angles should not be
taken for granted in pulsar population synthesis, and the sensitivity of
its results to this assumption should be tested.

\section*{Acknowledgements}

A.B. was supported by the Simons Foundation grant MP-SCMPS-00001470.

\appendix
\section{Results robustness}
\label{sec:app:robustness}

To find out to what extent the results of Section~\ref{sec:results}
reflect the properties of the pulsars rather than those of the
numerical scheme, we applied the scheme to mock data with known initial
parameters.

\subsection{A single pulsar}

First, a pulsar with given $P_0$, $\alpha_0$ and $b_0$ was evolved
forward by a direct numerical integration of
system~\eqref{eq:rotation_evolution}, which does not use
Equations~\eqref{eq:main_eq}--\eqref{eq:initial_period}. With the exact
age and obliquity, the backward solution recovers the initial values to
a relative accuracy of $\sim 10^{-12}$. With a realistic age error, the
result depends on how much the pulsar has spun down. For
$P_0 = 0.1$~s, $\alpha_0 = 30^\circ$, $\beta = 0.2$,
$t_\mathrm{m} = 10^5$~yr and an age of $10^{6.5}$~yr, the pulsar is
observed with $P = 0.37$~s and $\alpha = 10^\circ$, and
limit~\eqref{eq:thetamax} is reached at an age only $0.024$~dex above
the true one. Close to this limit $\alpha_0$ and $P_0$ depend on the
assumed age very steeply: ages of $10^{6.48}$, $10^{6.50}$ and
$10^{6.52}$~yr give $\alpha_0 = 25^\circ$, $30^\circ$ and $46^\circ$ and
$P_0 = 0.13$, $0.10$ and $0.045$~s. For an age error of $0.1$~dex about
40 percent of the age draws have no solution, and only the younger
ones survive. As a result, the reconstructed values are biased: the
median $\alpha_0$ is $\approx 20^\circ$ instead of $30^\circ$ and the
median $P_0$ is $0.16$~s instead of $0.10$~s. The bias vanishes when
the pulsar has spun down by less than a factor of $\sim 2$
($P/P_0 \lesssim 2$) and grows for larger $P/P_0$: initial periods much
shorter than the present one cannot be recovered, and the reconstructed
$P_0$ tends to a fixed fraction of $P$.

\subsection{A mock population}

Second, we generated mock samples of 35 pulsars from known parent
distributions: initial periods from the normal distribution of
\citet{fgk06} ($0.30 \pm 0.15$~s), isotropic initial magnetic angles, and
initial magnetic fields from a log-normal distribution with
$\langle\log B_0\,[\mathrm{G}]\rangle = 12.5$ and
$\sigma_{\log B_0} = 0.5$. The true ages were set equal to the
kinematic ages of the 35 pulsars of Table~\ref{tab:pulsars}. Each mock
pulsar was evolved forward to its age and accepted with the probability
$f(\alpha, P)$ of Equation~\eqref{eq:beaming}, so that the mock sample
is subject to the same geometric selection as the real one. The mock
kinematic ages were taken equal to the true ones (i.e. unbiased), with
the error bars of Table~\ref{tab:pulsars}. The samples were then
processed by the same scheme as the real data, with the same
(known) $\beta$ and $t_\mathrm{m}$.

\begin{figure*}[!t]
\centering
\paperfigure{0.9\textwidth}{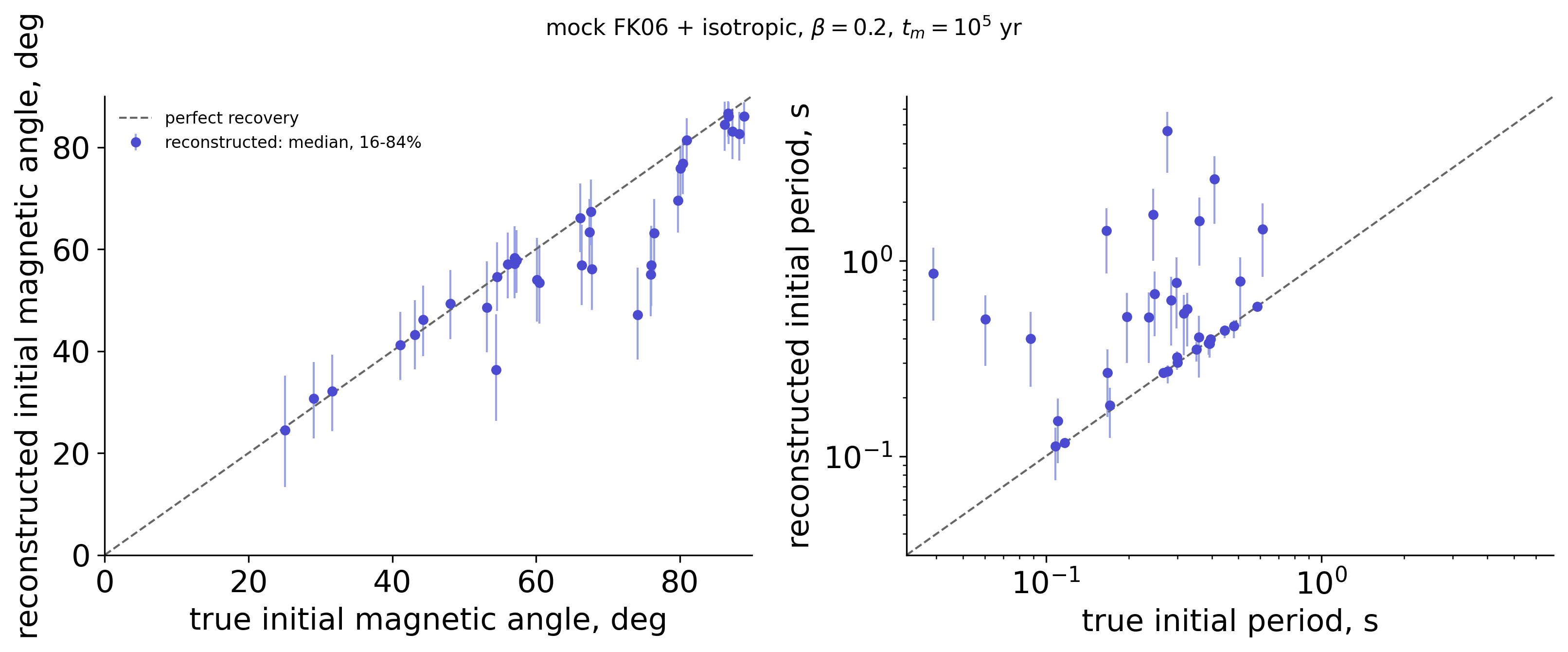}
\caption{
Reconstructed versus true initial magnetic angles (left) and initial
periods (right) of the 35 mock pulsars drawn from the FK06 distribution
of initial periods and the isotropic distribution of initial angles
($\beta=0.2$, $t_\mathrm{m}=10^5$~yr). The points and bars show the
median and the 16--84 percent range over the Monte Carlo trials.
}
\label{fig:pop_perpulsar}
\end{figure*}

\begin{figure*}[!t]
\centering
\paperfigure{\textwidth}{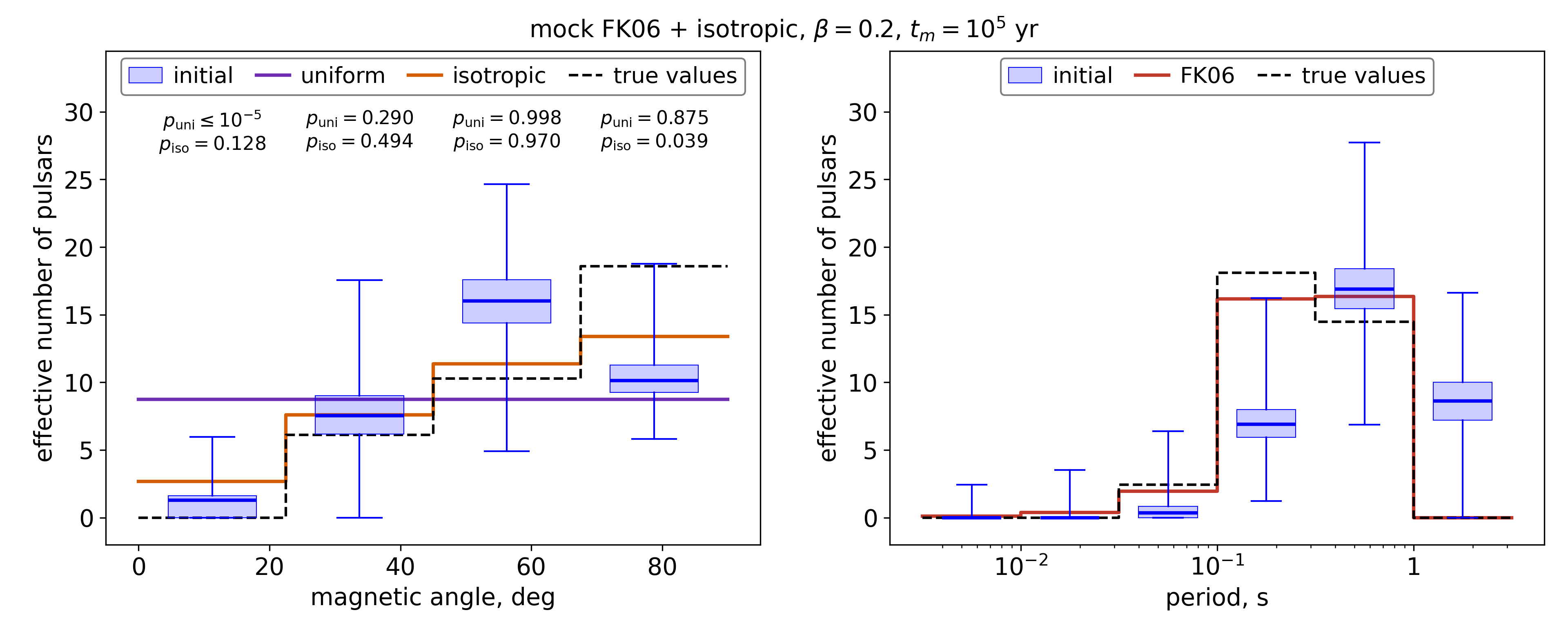}
\caption{
Reconstructed distributions of the initial magnetic angles (left) and
initial periods (right) for the same mock sample as in
Figure~\ref{fig:pop_perpulsar} (box plots), compared with the parent
distributions (red) and with the distributions of the true initial
values of the 35 mock pulsars (dashed).
}
\label{fig:pop_distr}
\end{figure*}

The results for $\beta = 0.2$, $t_\mathrm{m} = 10^5$~yr are shown in
Figures~\ref{fig:pop_perpulsar} and~\ref{fig:pop_distr}; for $\beta=0$
they are similar. The initial $\alpha$ of the individual pulsars is
recovered without a significant bias
(Figure~\ref{fig:pop_perpulsar}, left): the median difference between
the reconstructed and the true $\alpha_0$ is $-2^\circ$, and the true
value lies within the 16--84 percent interval of the reconstruction
for $71$--$74$ per cent of the pulsars.

The distribution of $\alpha_0$, however, is reproduced less accurately
than the individual values (Figure~\ref{fig:pop_distr}, left). The
smearing of $\alpha$ and the truncation of the ages move a part of the
pulsars from the last bin to the neighboring one: in the example shown,
the number of pulsars with $\alpha_0 > 67.5^\circ$ is $18.6$ for the
true values and $10.1^{+1.8}_{-1.3}$ after the reconstruction, and for
$\beta = 0$ these numbers are $10.8$ and $8.9^{+1.4}_{-1.3}$. As a
result, even for an isotropic parent distribution the reconstructed
number in the last bin falls below the isotropic expectation of
$13.4$, with $p_\mathrm{iso} = 0.04$ and $0.002$, respectively. The
method, therefore, produces by itself a moderate deficit of nearly
orthogonal rotators, and small values of $p_\mathrm{iso}$ alone do not
prove that the parent distribution is not isotropic.

Nevertheless, the deficit found for the real pulsars is stronger than
the one produced by the method. The real sample contains
$5.5^{+1.0}_{-0.8}$ ($\beta=0$) and $6.4^{+1.2}_{-1.0}$
($\beta = 0.2$) pulsars with $\alpha_0 > 67.5^\circ$, to be compared
with $8.9$ and $10.1$ for the mock samples, and the corresponding
probabilities are smaller by two to three orders of magnitude
($p_\mathrm{iso} \le 10^{-5}$ and $2\times10^{-5}$ against $0.002$ and
$0.04$). In an ensemble of 200 mock samples, the reconstructed number of pulsars in the
last bin is $8.8 \pm 2.9$ ($\beta=0$) and $9.7 \pm 2.7$
($\beta=0.2$), i.e. on average $3.5$--$4.5$ pulsars below the true one, and
it is as small as in the real sample in $\approx 9.5$ and $\approx 8.0$
per cent of the realizations. We conclude that the lack of pulsars with
large initial $\alpha_0$ is not a pure product of the numerical
scheme and more likely reflects a real property of the pulsars;
however, its statistical significance is moderate, and an isotropic
parent distribution cannot be rejected at a level better than
$\sim 90$ per cent.

The situation with the initial periods is different
(Figures~\ref{fig:pop_perpulsar} and \ref{fig:pop_distr}, right). In
agreement with the single-pulsar test, the periods of the mock pulsars
that have spun down considerably are overestimated, some of them by an
order of magnitude. The median of
$\log(P_{0,\mathrm{rec}}/P_{0,\mathrm{true}})$ is $+0.13$ to
$+0.19$~dex, and the true value lies within the 16--84 per cent
interval for only $54$--$60$ per cent of the pulsars. The
reconstructed distribution loses most of the pulsars with
$P_0 = 0.1$--$0.3$~s and acquires a spurious tail at $P_0 > 1$~s, where
no mock pulsar was born. For the FK06 parent distribution the scheme returns
$m_{P_0} = 0.9$--$1.0$~s and $m_{\log P_0} = -0.26$ and $-0.20$ for
$\beta = 0$ and $0.2$, respectively. These values are even larger than
those obtained in our analysis for the real pulsars ($m_{\log P_0} = -0.43$ and
$-0.52$). Therefore, the long initial periods found in
Section~\ref{sec:res_p0} do not contradict the distribution of
\citet{fgk06}: they are what the method returns for a population with
shorter periods, and the real initial periods are probably shorter than
the reconstructed ones. A caveat is that the mock samples were not
subject to any selection by the radio luminosity or by the death line
and contain pulsars with present-day periods of up to $7$--$8$~s, which
are absent in the real sample; this probably exaggerates the bias. An
unbiased estimate of the initial periods requires a forward modelling
of the sample, which is beyond the scope of this paper.

Two more limitations should be mentioned. The tests assume that the
spin-down law and the field decay are known exactly and that the
kinematic ages are unbiased, whereas the large fraction of pulsars with
$t_\mathrm{kin} > \tau_\mathrm{ch}$
(Section~\ref{sec:compatibility}) shows that at least one of these
assumptions is violated for a part of the sample. And the errors of the
observed magnetic angles may be larger than the adopted $\pm10^\circ$,
especially at large $\alpha$.

\bibliographystyle{elsarticle-harv}
\bibliography{eman}

\end{document}